%% file: culprit_arxiv.tex
\documentclass{article}

     \PassOptionsToPackage{numbers, compress}{natbib}

     \usepackage[preprint]{neurips_2019}

\usepackage[utf8]{inputenc} 
\usepackage[T1]{fontenc}    
\usepackage{hyperref}       
\usepackage{url}            
\usepackage{booktabs}       
\usepackage{amsmath,amssymb,amsthm}
\usepackage{bm}
\usepackage{amsfonts}       
\usepackage{nicefrac}       
\usepackage{microtype}      
\usepackage{subcaption}
\usepackage{wrapfig}
\usepackage{multirow}

\usepackage{algorithm}
\usepackage{algpseudocode}

\title{Approximate Identification of the Optimal Epidemic Source in Complex Networks}

\author{%
  S. Jalil Kazemitabar \\
  Department of Statistics\\
  University of California, Los Angeles \\
  \texttt{sjalilk@ucla.edu} \\
   \And
   Arash A. Amini \\
   Department of Statistics \\
   University of California, Los Angeles \\
   \texttt{aaamini@ucla.edu} \\
}

\input{defs}

\usepackage{arash_macros}

\usepackage[inline]{enumitem}

\begin{document}

\maketitle

\begin{abstract}
  We consider the problem of identifying the source of an epidemic, spreading through a network, 
  from a complete observation of the infected nodes in a snapshot of the network. 
  Previous work on the problem  has often employed geometric, spectral or heuristic approaches to identify the source, 
  with the trees being the most studied network topology. We take a fully statistical approach and derive novel recursions to compute the Bayes optimal solution, under a susceptible-infected (SI) epidemic model.  Our analysis is time and rate independent, and holds for general network topologies. We then provide two tractable algorithms for solving these recursions, a mean-field approximation and a greedy approach, and evaluate their performance on real and synthetic networks. Real networks are far from tree-like and an emphasis will be given to networks with high transitivity, such as social networks and those with communities. We show that on such networks, our approaches significantly outperform geometric and spectral centrality measures, most of which perform no better than random guessing. 
  Both the greedy and mean-field approximation are scalable to large sparse networks. 
  
\end{abstract}

\section{Introduction}
Modern transportation networks have had profound effects on geographical spread of infectious diseases~\cite{cliff2004,cohen2000} giving rise to complicated epidemic evolutions~\cite{colizza2006role}. These evolutions can be modeled as dynamic processes on transportation networks. The epidemic spread on networks can take other forms, such as outbreaks of foodborne diseases~\cite{slutsker1998foodborne}, intercontinental cascade of failures among financial institutions~\cite{elliott2014financial,acemoglu2015systemic}, computer malware propagation on the interent and mobile networks~\cite{kondakci2008epidemic,fleizach2007can} 
spread of targeted  fake news~\cite{shao2017spread,shao2018spread} and rumors~\cite{friggeri2014rumor} on social media,
 especially during presidential elections~\cite{shin2017political,jin2017detection,allcott2017social}.
In response to an adverse diffusion on a network, it is critical to trace back sources  to enable appropriate prevention and containment of the spread~\cite{world2008foodborne}. 
 Inferential methods have been developed to locate the source of foodborne diseases~\cite{manitz2014,horn2019}
 and influenza pandemics~\cite{neumann2009,shen2016}.
   In the context of online social networks, 
   the spread of misinformation can be limited by the identification of influential users~\cite{pei2014,kitsak2010}. Source recovery can also be used to assess the power of diffusions in generating anonymity in network protocols~\cite{bojja2017}. 

%
%

The epidemic source identification problem has received considerable attention in the past decade. Given a snapshot of the infected nodes in a network, the task is to discover who has originated the epidemics. 
Since the seminal work of Shah and Zaman~\cite{Shah2011}, 
numerous attempts have been made
to address the question and its extensions~\cite{Fioriti2012,Lokhov2014,Zhu2016,Luo2012,Nguyen2016,Prakash2012}.
By now, there are multiple methods that show satisfactory results in limited experimental setups or have proven guarantees in restricted network topologies~\cite{Jiang2017}. However, identifying the source under general conditions still remains a difficult task. Even under fairly simple models such as the Independent Cascade (IC) dynamics, the problem of optimal recovery appears to be NP-hard in infection size~\cite{Lappas2010}. The theoretical guarantees for optimal 
and consistent recovery are restricted to regular infinite trees \cite{Shah2011, Zhu2016}, and as we show in this paper, the popular and well-cited methods are quite unreliable in a wide range of real networks.

Source identification has remained largely unsolved and poorly understood for real complex networks. As we will show  through experiments in Section~\ref{sec:sims}, in real networks, even the optimal Bayes estimator applied to small infected sets has difficulty narrowing down to the true source.
It is thus important to recover as much information from the likelihood of the model as possible.
We develop techniques for computing the full likelihood of the infection, as opposed to identifying the most likely sample-path~\cite{Zhu2016}.
Moreover, we fully exploit the information from the boundary of the infection set, in addition to the structure inside the infected subgraph. This idea has been pointed out before~\cite{Chang2015}, but has been mostly neglected by subsequent work; cf.~\cite{Prakash2012, Fioriti2012}. We develop all these ideas without restricting the structure of the network to trees. Our framework also easily extends to the case where there are multiple infecting sources (Appendix~\ref{sec:multi:source}). 

In this paper, we develop statistical algorithms that outperform the state-of-the-art 
in a wide range of network topologies. Our contributions are distinct in several ways:
\vspace{-1ex}
\begin{enumerate}[leftmargin=*,itemsep=0pt]
	\item Our methods are parameter-free, meaning that they do not require knowing the duration of the epidemic or how fast it grows.
	\item We show that the exact maximum likelihood estimator (MLE) of the source---or equivalently the Bayes optimal solution under uniform prior---can be written as a dynamic programming (DP), with easily computable coefficients based on the adjacency matrix of the network.
	\item We develop two schemes to approximate the DP: an efficient greedy elimination (GE), and a novel mean-field approximation (MFA) of the likelihood,  computed by solving a linear system. 
	\item Our approximations are more disciplined than existing approaches. 
	They do not impose restrictions on the topology of the network. Nor do they appeal to the partial likelihood of the candidate infecting sets. This is in contrast to the use of 
	spanning trees to deal with general topologies~\cite{Shah2011} or the \emph{path-based approaches} that rely on the likelihood of individual paths from potential sources to the infected set~\cite{Zhu2016}. 
\end{enumerate}
\vspace{-1ex}

We will show that when applied to real networks, both approximation schemes (MFA and GE) 
outperform various geometric and spectral approaches, most of which perform no better than random guessing.  We also show that even for basic models of real networks, e.g., models with community structure, most existing methods dramatically fail.
%
%
The improvement in performance is most significant for the networks with many cycles, including social networks 
that are known to have high transitivity. 
In terms of computational efficiency, both the greedy and mean-field approximations are superior to the state-of-the-art likelihood-based and spectral approaches and comparable to centrality-based methods.
%
In addition, the mean-field algorithm is easily parallelizable through standard linear algebraic routines and can be used to tackle very large-scale epidemics on real networks.


\paragraph{Related work.}
Most of the existing literature on the source identification problem are based on a SIR dynamic where the infection spreads with an exponential rate proportional to the number of infected neighbors. All nodes are \emph{susceptible} to the infection and once \emph{infected} may \emph{recover} with a fixed exponential rate~\cite{kiss2017}. Moreover, the spread of infection through edges are mutually independent. Different variations of SIR may assume that no recovery is possible (SI) or the recovered is not immune to iterated infections (SIS).


Shah and Zaman~\cite{Shah2011} considered the SI dynamics and proposed the Rumor Centrality (RC), which counts the \emph{permitted permutations}, a.k.a. infection paths, inside the infected subgraph. 
Their linear time algorithm is an optimal estimator in regular trees and enjoys strong theoretical properties in such idealized settings~\cite{khim2017}. 
%
Zhou and Ying \cite{Zhu2016} consider SIR dynamics on a tree and show that the most likely infection path is rooted at a Jordan center (JC) of the infected set $O$, that is, a node with minimum eccentricity (i.e., maximum distance to others). It has been shown~\cite{Zhu2016,khim2017} that in regular trees, eccentricity ranking generates, with high probability, a confidence set containing the true source, whose size does not grow with the infection size.
%

The Dynamic Message Passing (DMP) was proposed in~\cite{Lokhov2014} as an approximation of the maximum likelihood estimator in discrete SIR epidemics, by assuming that  the marginal probabilities of infection for each node are independent.
Despite compelling performance, DMP 
is computationally intensive and impractical for large networks with moderately dense structures, even for small infection sets.
A spectral algorithm, called Dynamical Age (DA) was introduced in~\cite{Fioriti2012}, based on measuring the sensitivity of the maximum eigenvalue of the Laplacian matrix to the elimination of each node in the infection set. The algorithm was mainly developed to discover the initial node in a growing preferential attachment model. Another spectral method for the discrete SI model is proposed in~\cite{Prakash2012}.

\section{Source detection in SI epidemics}
We consider a continuous-time susceptible-infected (SI) epidemic~\cite{kiss2017}
with rate of infection~$\beta$, on a static undirected network $G(V,E)$ with known edge set $E$ and $V = [n]$. At time zero, all nodes but the source  are in the  susceptible state. Infection is a terminal state and 
susceptible nodes 
are exposed to the infection at an exponential rate proportional to the number of their infected neighbors. More precisely, given that nodes $I$ are infected at some time $t$, we run exponential clocks $T_j \sim \Expd( \beta \vol(I,j))$ for all $j \in I^c$ and the first to expire determines the next infected node: If $\bm j^* = \argmin_{j} T_j$, then the dynamics move to the infected set $I \cup \{\bm j^*\}$ at time $t + T_{\bm j*}$.
It is clear that the contagion will eventually spread through the entire graph.

The infection source or patient zero, denoted as $\is$, is unknown. What we observe is a snapshot of the contagion at some time $t$, meaning the entire set of infected nodes at that time, which we denote by $O$.
The objective is to find $\is \in O$ or form a confidence set for $\is$ with desired false exclusion probability. Our focus here will be on the single source setting, but the analysis 
is extensible to the multi-source setting
(cf. Section~\ref{sec:multi:source}). 

\textbf{Notation.} 
We write $A \in \{0,1\}^{n \times n}$ for the adjacency matrix of the network 
%
and  $\vol(I,J) := \sum_{i \in I, j \in J} A_{ij}$ for the volume of a cut in the network between subsets $I, J \subset [n]$ of nodes. For singleton subsets,
we often drop the braces, 
e.g., $\vol(I, j) := \vol(I, \{j\})$ and $O \setminus j = O \setminus \{j\}$.

\subsection{Time and rate invariant analysis}
We start by examining the  probability of observing a particular set of infected nodes given a starting source. 
Let us introduce a parameter-free formulation of the problem (i.e. not dependent on rate $\beta$ and time $t$) that will be the foundation for our analysis of the continuous SI dynamics. The idea has been introduced in~\cite{Chang2015}. We generalize it to multiple sources and find 
forward and backward dynamic programming
formulations that allow us to derive efficient approximations.

Suppose that, at some point in time, the infection reaches $I\subset [n]$. Let $O \subset [n]$ be some superset of $I$. We are interested in computing $\rho_{I \to O}$, the chance that all the nodes in $O$ are infected before any node outside. More precisely, let
\begin{align}\label{eq:def:rho}
\rho_{I \to O} := \pr\big( \text{$O$ is infected before $O^c$} \mid \text{$I$ is infected}\big).
\end{align}
We refer to $\rho_{I \to O}$ as the \emph{transition probabilities}.
Note that these transition probabilities are independent of the infection source.
Given that in a snapshot of the contagion, nodes $I$ are infected,
$\rho_{I \to O}$ determines how likely it is that in some future snapshot, $O$ is the set of infected nodes. The Markov property of (continuous-time) SI dynamics allows us to define $\rho_{I \to O}$ without reference to the source, or the time of the first snapshot. We will also show that these probabilities do not depend on the infection rate or the time of the second snapshot. 

\subsection{Statistical Inference}\label{sec:bayes:infer}
%

Given the observed (random) infected set $O$,  the function $I \mapsto \rho_{I \to O}$ is \emph{the likelihood} of the model. Writing $ L_{O}(I) := \rho_{I \to O}$ for this likelihood, we observe that $L_O(I) = 0$ for all $I$ not contained in $O$. So, we can restrict $L(\cdot)$ to all subsets of $O$. 
When dealing with the single-source setup, we restrict the parameter space to $I = \{i\}$ and with some abuse of notation write $\rho_{i\to O}$ for $\rho_{\{i\} \to O}$, and $ L_O(i) = \rho_{i \to O}, i \in [n]$ for the likelihood.

We can further consider a Bayesian setup by putting a uniform prior on the source (i.e., uniform over $[n]$). The Bayesian setup allows us to consider various notions of optimality by changing the loss function. 
%
%
Letting $\bm i_*$ be the random initial source, we have a joint distribution on $(\bm i_*, O)$. Then the posterior probability that the source is $i$, given that we observed infected nodes $O$ is
\begin{align*}
p_i := \pr (\bm i_*  = i\mid O) = \frac{\rho_{i \to O}}{\sum_{j\in O} \rho_{j \to O}} 1\{ i \in O\}.
\end{align*}
Therefore, the maximum a posteriori (MAP) estimate of the source is
$\bm i^*_\text{MAP} = \argmax_{i}\, \rho_{i \to O}$
which minimizes the probability of error. 
That is, $i^*_{\text{MAP}}$ minimizes $\pr(\ih \neq \bm i_*)$ for any estimator $\ih =  \ih(O)$. 
In some applications, the graph geodesic distance ($d_G$) to the source determines the error of estimation. In that case, the Bayes optimal estimator is
$\bm i^*_\text{dist} = \argmin_{i} \sum_{j\in O} \text{dist}_G(i, j) \, \rho_{j \to O}.$
It is not hard to see that $\bm i^*_\text{dist}$ minimizes $\ex [d_G(\ih, \bm i_*)]$ among all possible estimators $\ih$.

A third choice is to output a ranking instead of a single source. In this case an estimator is formally a permutation $\sigh = \sigh_O$ on $[n]$, suppressing the dependence on $O$ for simplicity. We can then consider the \emph{rank loss}  $\ell(\sigh, \bm i_*) = \sigh(\is)$, and we call the associated risk the \emph{expected (source) rank} $= \ex  \sigh(\is)$.
%
The corresponding optimal Bayes estimator is obtained by minimizing the  posterior risk:
\begin{align*}
\sigh^* := \argmin_{\sigma: [n] \to [n]}\; \ex [ \sigma(\is)\mid O].
\end{align*}
Noting that $\ex [ \sigma(\is)\mid O] = \sum_i \sigma(i) \,p_i$, the optimal estimator in this case is the ranking that sorts $p_i$ into descending order, i.e., $\sigh^*(j_i) = i$ where $p_{j_1} \ge p_{j_2} \ge \cdots \ge p_{j_n}$.


\begin{rem}
	The distance loss might be suitable in some applications, but in general it is a poor measure if the goal is to reveal the actual source. This is especially true in small world networks,  including most social networks, where the expected distance between any pair of nodes is small. On the other extreme, in terms of the precision in recovering the source, is the zero-one loss which is too stringent. The rank loss can be considered a more robust version of the 
	zero-one loss, and we will be our main evaluation measure.
	
\end{rem}

\section{Exact likelihood computation}
The Bayesian estimators introduced in Section~\ref{sec:bayes:infer} require us to evaluate the posterior probabilities $(p_i)$, or equivalently the likelihood values $\rho_{j \to O}$ for all $j\in O$. The main difficulty of the source identification problem is that computing the likelihood is itself challenging. We now develop exact equations that allow us to recursively compute the likelihood values $L_O(I)$ for all subsets $I \subset O$. 

\paragraph{Dynamic programming.}
To begin, note that $\rho_{O \to O}=1$ for any $O \subset [n]$. In addition,  $\rho_{I \to O}=1$ whenever $O$ corresponds to a connected component of $G$. We develop two dynamic programming expressions for $\rho_{I \to O}$ for general $I\subset O$:
\begin{prop}\label{prop:bayes}
	For $I \subset O \subset [n]$, the probabilities $\rho_{I \to J}$ defined in~\eqref{eq:def:rho} satisfy the forward program 
	\begin{align}\label{eq:for:rec}
	\rho_{I \to O} = \sum_{j\in O\setminus I} \frac{\vol(I, j)}{\vol(I, I^c)}\, \rho_{I\cup j \to O}
	\end{align}
	and the backward program
	\begin{align}\label{eq:back:rec}
	\rho_{I \to O} = \sum_{j\in O\setminus I} \rho_{I \to O\setminus j} \,\frac{\vol(O\setminus j, j)}{\vol(O \setminus j, (O \setminus j)^c)}.
	\end{align}
\end{prop}

In the forward programming~\eqref{eq:for:rec}, $j$ effectively iterates over the boundary of $I$ in $O$, as $\vol(I,j)=0$ if $j$ is outside that boundary. Therefore, the running time of the forward programming benefits from the sparsity of the network. Unlike the forward programming, the iteration over $j$ in~\eqref{eq:back:rec} cannot be restricted to a smaller set. 
A corollary of Proposition~\ref{prop:bayes} is that the transition probabilities $\rho_{I \to J}$ are not affected by the rate and the duration of the infection. 


Let us now observe some connection with the \emph{path-based analysis}. A permitted permutation or an infection path starting at a node $\is$, refers to a permutation $\sigma$ of nodes with $\sigma_1= \is$, and such that $\sigma_{k+1}$ is connected to at least one node in $\{\sigma_1,\ldots,\sigma_k\}$, for all $k\in [|\sigma|-1]$. 
 Notice that the probability of observing a given infection path is 
\begin{align}\label{eq:infect:path:prob}
\pr \big( \text{path $\sigma$ observed} \mid \sigma_1=\is \big) = \prod_{k=1}^{|\sigma|-1} \frac{\vol(\sigma_{[k]}^{}, \sigma_{k+1})}{\vol(\sigma_{[k]}^{}, \sigma_{[k]}^c)}
\end{align}
where $\sigma_I := (\sigma_i \mid i \in I)$. 
As noted in~\cite{Chang2015}, one can obtain the transition probability $\rho_{\{\is\} \to O}$
by summing~\eqref{eq:infect:path:prob} over all infection paths $\sigma$ such that $\sigma_1=\is$ and $\{\sigma_1,\ldots,\sigma_k\}=O$. 
 Our recursive representation is novel, avoids these explicit summations, and will be key  in deriving  approximation schemes for $\rho_{I \to O}$ in Section~\ref{sec:approx}.

Path-based approaches such as Jordan center~\cite{Zhu2016} forgo computing the  complete likelihood (i.e., avoid summing the odds of all infection paths) and instead 
find the most probable path, that is, one that maximizes~\eqref{eq:infect:path:prob}. In contrast, equations~\eqref{eq:for:rec} and~\eqref{eq:back:rec} compute the complete likelihood of the infection set, which has the following advantages over the path-based likelihood: 
 It fully exploits the structure of the graph inside the infection set, not just a spanning tree or a permitted permutation of nodes in the infected subgraph. Moreover, it takes into account the boundary of the infected subgraph via $\vol(I, I^c)$.


\textbf{NP-hardness.}
The recursions we derived are a dynamic programming (DP) solution that computes the likelihood more efficiently, by avoiding the direct summation over all possible paths. 
More precisely, one can obtain the optimal likelihood values by solving recursion~\eqref{eq:for:rec} backwards: Starting from $\rho_{O\to O} = 1$, we can determine $\rho_{I\to O}$ for all proper subsets $I \subset O$ of maximal size (i.e., $|I| = |O|-1$) and then all the proper subsets of those at the previous stage and so on. 
%
However, this DP procedure  may still take $O(2^{|O|})$ time in the worst case.  
It is an interesting open question whether polynomial-time solutions for computing the likelihood or its maximum exist.

Although it has been shown in~\cite{Shah2011} that for infinite regular trees, computing the likelihood reduces to enumerating all the infection paths from $i$ to $O$, and therefore a polynomial-time algorithm exists in that case, the behavior of the exact likelihood under more general network topologies is much more difficult to investigate. Researchers have implicitly assumed the problem to be hard beyond trees and have resorted to approximations that are often based on heuristics. In the next section, we propose disciplined approximation schemes that yield much more accurate results under realistic classes of networks.

\section{Approximations}\label{sec:approx}
We now provide two approximations to the likelihood function $L_O(I)$ based on the exact dynamic programming developed in Proposition~\ref{prop:bayes}.

\paragraph{Greedy Elimination (GE).}
We can obtain a singleton source set $I=\{i\}$ that maximizes $\rho_{I \to O}$ with greedy elimination of elements in $O$. The algorithm we propose is based on the backward recursion~\eqref{eq:back:rec} and is detailed in Algorithm~\ref{alg:greedy}. We start with $O_0:=O$ and consider all maximal proper subsets of $O_0$ that induce a connected subgraph of $G$. Among those, we choose the one that maximizes the transition probability to $O_0$, i.e. $\rho_{O_0\setminus j \to O_0}=\vol(O_0\setminus j, j) / \vol(O_0\setminus j, (O_0\setminus j)^c)$. Suppose that $O_1:=O_0\setminus j^*$ is the maximizer. Next, we iterate the same procedure for $O_1$ and so forth, until we reach a singleton set $I:=O_{|O|-1}$. 
The procedure has an $O(k^2 m)$ runtime where $k = |O|$and $m$ is the number of edges in the infected subgraph, $G_O$.

GE has a Bayesian justification. Let $\widetilde{O}_k$ be  the random infected set after $k$ steps. Suppose that we want to find the MAP for $\widetilde{O}_{k-1}$ given $\widetilde{O}_k$. The Bayesian posterior probability is
\begin{align*}
\pr(\widetilde{O}_{k-1}=O\backslash j \,|\, \widetilde{O}_k=O) \propto \rho_{O\backslash j \to O} \cdot P(\widetilde{O}_{k-1}=O\backslash j).
\end{align*}
Whenever $G_{O\backslash j}$ is connected, the prior is positive. GE finds a proxy for MAP through maximizing the evidence and ensuring the prior is positive.

Algorithm~\ref{alg:greedy} has similarities with finding the most likely path from a source to the observed snapshot. Chang et. al.~\cite{Chang2015} propose a similar path-based search called GSBA. They start from each node in $O$ and approximate the most likely path and use it as a proxy to the most likely source. Algorithm~\ref{alg:greedy}, however, does this greedy search in a backward fashion. 

\begin{figure}[t]
	\begin{minipage}[c]{0.53\textwidth}
		\vspace{0pt} 
		\begin{algorithm}[H]
			\begin{algorithmic}[1]
				\Require {Graph $G([n],E)$ and  $O\subset [n]$.} 
				\Ensure {$\bm i_\text{GE}^* \in O$}.
				\State $O_0 := O$
				\For {$i := 0$ to $|O|-2$}
				\State $O'_i := \{j\in O_i : G_{O_i\backslash j}\text{ remains connected} \}$
				\State $j^* := \argmax_{j\in O'_i}  \dfrac{\vol(O_i\setminus j, j)}{\vol(O_i\setminus j, (O_i\setminus j)^c)}$.
				\State $O_{i+1} := O_i \setminus j^*$.
				\EndFor
				\State {$\bm i_\text{GE}^* :=$ the single element in $O_{|O|-1}$}.
			\end{algorithmic}
			\caption{Greedy Elimination}
			\label{alg:greedy}
		\end{algorithm}
	\end{minipage}
	\hfill
	\begin{minipage}[c]{0.45\textwidth}
		\vspace{0pt} 
		\begin{algorithm}[H]
			\caption{Mean-Field Approximation}
			\begin{algorithmic}[1]
				\Require {Graph $G([n],E)$ and  $O\subset [n]$.} 
				\Ensure {$\bm i^*_\text{MFA} \in O$}.
				\State Compute $S, \zb$ as defined in~\eqref{eq:Sz:def}.
				\State $\bbh := S^{-1}\zb$.
				\State $\bm i^*_\text{MFA} := \argmax_{j\in O} \bbh_j$.
			\end{algorithmic}
			\label{alg:mean:field}
		\end{algorithm}
	\end{minipage}
\end{figure}

\paragraph{Mean-field Approximation (MFA).}
We now approximate $\rho_{I \to O}$ by the mean-field technique. The idea is to treat the set function $I \mapsto \rho_{I \to O}$ as if it was a distribution (or measure) on $O$ and approximate it by the product of its marginals.
Fix a subset $O\subset [n]$. For any $I\subset O$, let $\xb^I = (x^I_j)_{j\in O}$ be the binary representation of $I$, i.e. $x^I_j = 1\{j \in I\}$ for any $j\in O$. We find 
$\alpha_0$ and $(b_j)_{j\in O}$ such that 
\begin{align}\label{eq:rho:hat:mean:field}
\widehat{\rho}_{I \to O} = \alpha_0 \prod_{j \in O} b_j^{x^I_j-1}
\end{align} 
is a good approximation to $\rho_{I \to O}$ for all $I \subset O$, in the sense of minimizing the quadratic deviation from the solution of the recursion~\eqref{eq:for:rec}. 
First note that $\alpha_0=1$ since $\rho_{O \to O} = 1$.
Next, we plug-in $\widehat{\rho}_{I \to O}$ into the forward recursion, to get
\begin{align*}
\vol(I, I^c)\,\widehat{\rho}_{I \to O} - \sum_{j \in O\setminus I} \vol(I,j)\, \widehat{\rho}_{I\cup \{j\} \to O} = 0.
\end{align*}
Dividing both sides by $\prod_{j\in O\setminus I} b_j$ gives
$\vol(I, I^c) - \sum_{j \in O\setminus I} \vol(I,j)\, b_j = 0.$
These equations in general cannot be satisfied exactly for all $I \subset O$. Instead, letting $\bb = (b_j)_{j \in O}$, we solve the following least-squares problem:
\begin{align}\label{eq:optim:mean:field}
\bbh \;\in\; \argmin_{\bb} 
 \sum_{I:\,I\subset O} \Big(\vol(I, I^c) - \sum_{j \in O\setminus I} \vol(I,j) \,b_j\Big)^2 = \argmin_{\bb} \norm{Q \bb - \rb}_2^2
\end{align}
where $Q \in \reals^{(2^{|O|}-1) \times |O|}$ and  $\rb \in \reals^{(2^{|O|}-1) \times 1}$  are defined as follows:
\begin{align*}
Q_{I, j} &= 1\{j\not\in I\}\, \vol(I,j),\; \forall\, I\subset O,\; j\in O, \qquad 
\rb_I = \vol(I, I^c),\; \forall\, I\subset O.
\end{align*}
The solution of~\eqref{eq:optim:mean:field}  satisfies the normal equations $Q^T Q \bbh = Q^T \rb$. 
%
The following proposition shows that $Q^T Q$ and $Q^T \rb$ can be computed efficiently. Let $A$ be the adjacency matrix of the network.

\begin{prop}\label{prop:mean:field:lin:system}
	The solution $\bbh$ of~\eqref{eq:optim:mean:field} satisfies 
	the linear system 
	$S \bbh = \zb$
	with $S$  and $\zb$ given by
	\begin{align}
	\begin{split}
	S &= \Xi\big(A_{OO} + A_{OO}^2 -  
	A_{OO} \odot (\ub \one^T + \one \ub^T ) + \ub \ub^T \big) \in \reals^{|O| \times |O|}, \\
	\zb &= ( \one^T \ub + 2\one^T \vb) \ub  - 
	2 \vb \odot \ub + 2 (A_{OO} \,\vb + \ub) \label{eq:Sz:def}
	\end{split}
	\end{align}
	
	where $\ub := A_{OO}\one$ and $\vb = A_{OO^c}\one$. Here $\odot$ is the element-wise matrix product,  $\Xi(\cdot)$ is a matrix operator that returns the same matrix with double the diagonal entries, and $\one$ is the vector of all ones.
\end{prop}
See Appendix~\ref{app:proof:mean:field} for the proof. 
%
Proposition~\ref{prop:mean:field:lin:system} shows that the mean-field approach reduces to solving a linear system of equations in $|O|$ variables, a task with much better computational complexity than solving the original recursion.
Both $S$ and $\zb$ can be computed in at most $O(|O|^2)$ time. In the cases where $A$ is sparse (which often the case for real networks), $S$ will be a rank-one perturbation of a sparse matrix (both $A_{OO}$ and $A_{OO}^2$ will be sparse), hence solving the resulting system is often much faster than the worst-case, i.e., faster than $O(|O|^3$).


\section{Simulations}\label{sec:sims}
The methods proposed in this paper, the Greedy Elimination (GE) and the  Mean Field Approximation (MFA), have reasonable runtimes compared to popular source identification procedures and show superior performance in source identification. In this section, we make a comparison based on these two measures 
on real and synthetic networks.
As discussed in Section~\ref{sec:bayes:infer}, we consider ranking estimators (i.e., those outputting a permutation of the nodes according to their likelihood of being the source) and focus on the rank loss.
If the method does not return a ranking, we tweak it to do so. We evaluate the methods based on expected rank $R$, the expectation of the rank of the actual source among the list of candidates (cf. Section~\ref{sec:bayes:infer}). We normalize to get a number in [0,1], with zero corresponding to perfect recovery, i.e., we use $(R-1)/n$. 


\begin{figure}[t]
\centering
\begin{subfigure}[b]{0.45\textwidth}
\includegraphics[trim={1.1cm .6cm 1.1cm 1.2cm}, clip, width=\textwidth]{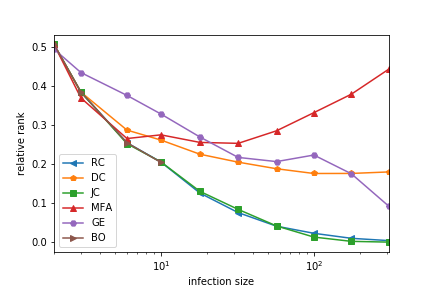}
\caption{Regular tree}
\label{fig:regular:tree}
\end{subfigure}
\begin{subfigure}[b]{0.45\textwidth}
\includegraphics[trim={0.85cm .6cm 1.1cm 1.2cm}, clip, width=\textwidth]{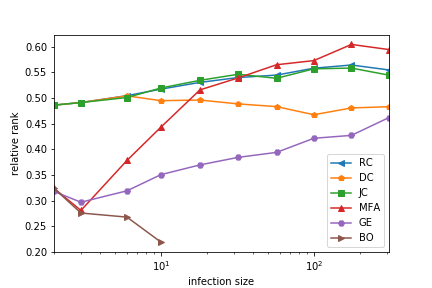}
\caption{Random tree}
\label{fig:random:tree}
\end{subfigure}
\begin{subfigure}[b]{0.45\textwidth}
\includegraphics[trim={1.1cm .6cm 1.1cm 1.2cm}, clip, width=\textwidth]{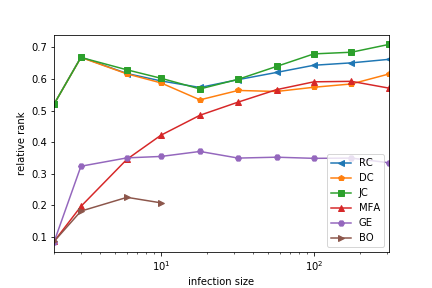}
\caption{Internet AS}
\label{fig:internet}
\end{subfigure}
\begin{subfigure}[b]{0.45\textwidth}
\includegraphics[trim={0.85cm .6cm 1.1cm 1.2cm}, clip, width=\textwidth]{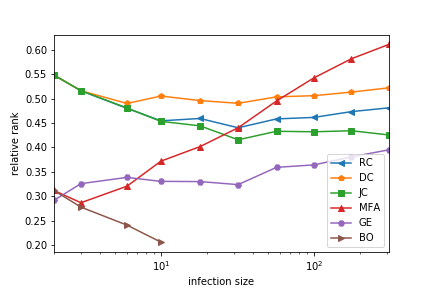}
\caption{US West Power Grid}
\label{fig:power}
\end{subfigure}
\caption{Plots of the expected relative rank versus the infection size for low-transitivity networks. }
\label{fig:rank:1}
\end{figure}

\begin{wraptable}{r}{.46\textwidth}
	\caption{Network statistics} 
	\label{tab:stats}
	\centering
	\renewcommand\arraystretch{1.2}
	
	\begin{tabular}{lcccc}
		\multirow{2}{*}{Network} &  \multirow{2}{*}{$n$} & mean & max & clust. \\
		&      & deg. & deg. &  coeff. \\
		\hline 
		Internet & 10670 & 4.0 & 2312 & 0.01 \\
		Power & 4941 & 3.0 & 19 & 0.1 \\
		Wiki vote & 7066 & 29.0 & 1065 & 0.13 \\
		UCSC68 & 8979 & 50.0 & 454 & 0.17 \\
		UC64 & 6810 & 46.0 & 660 & 0.19 \\
		DC-SBM & 1962 & 66.0 & 897 & 0.3 
	\end{tabular}
\end{wraptable}
We consider a variety of real and simulated networks.
Our selection includes an Internet Autonomous System~\cite{internetas,leskovec2005graphs}, US west-coast power grid~\cite{watts1998collective}, two Facebook-100 networks~\cite{traud2011comparing,traud2012social}, called UC64 and UCSC68, and a Wikipedia voting network~\cite{leskovec2010signed}. In addition, we present our results on a number of synthetic networks that are well studied in the literature, including regular trees, random trees, and degree-correlated stochastic block models (DC-SBM)~\cite{karrer2011stochastic}.

%
Table~\ref{tab:stats} summarizes the statistics on the largest connected component of these networks. The regular tree is of degree 3 and depth 10. The random tree has 500 nodes. For the DC-SBM network, we generate from a $3$-community planted partition version, i.e., $\ex[ A_{ij}] = \theta_i \theta_j P_{ij}$ where $P_{ij} = 0.5$ if nodes $i$ and $j$ are in the same community and $P_{ij} = 0.02$ if they are in different communities. The degree parameters $\theta_i$ are generated from a rescaled Pareto distribution with $\alpha =2$ and threshold $=1$.

\begin{figure}[t]
\centering
\begin{subfigure}[b]{0.45\textwidth}
\includegraphics[trim={0.85cm .6cm 1.1cm 1.2cm}, clip, width=\textwidth]{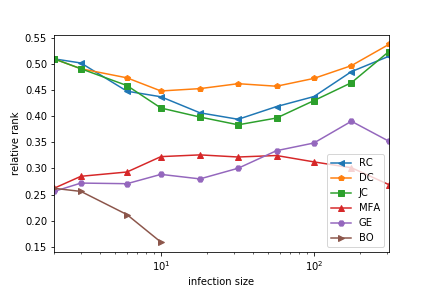}
\caption{UCSC68}
\label{fig:ucsc68}
\end{subfigure}
\begin{subfigure}[b]{0.45\textwidth}
	\includegraphics[trim={1.1cm .6cm 1.1cm 1.2cm}, clip, width=\textwidth]{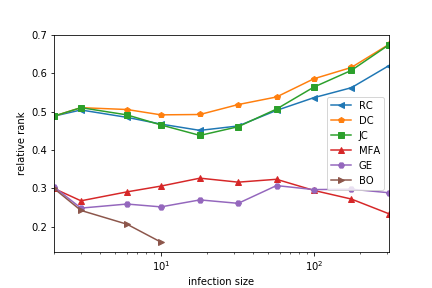}
	\caption{UC64}
	\label{fig:uc64}
\end{subfigure}
\begin{subfigure}[b]{0.45\textwidth}
	\includegraphics[trim={1.1cm .6cm 1.1cm 1.2cm}, clip, width=\textwidth]{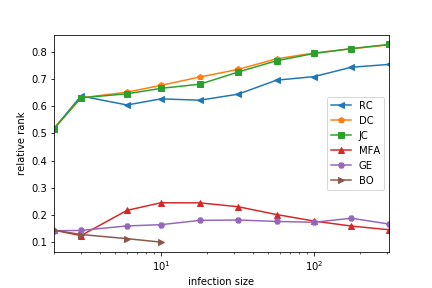}
	\caption{Wiki vote}
	\label{fig:wiki:vote}
\end{subfigure}
\begin{subfigure}[b]{0.45\textwidth}
\includegraphics[trim={1.1cm .6cm 1.1cm 1.2cm}, clip, width=\textwidth]{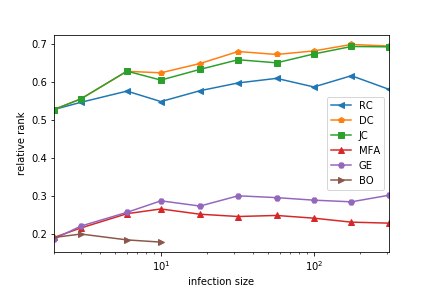}
\caption{DC-SBM}
\label{fig:dc:sbm}
\end{subfigure}
\caption{ Plots of the expected relative rank versus the infection size for  high-transitivity networks.}
\label{fig:rank:2}
\end{figure}

The results 
are illustrated in Figures~\ref{fig:rank:1}, \ref{fig:rank:2}. The methods we consider besides the optimal Bayes solution (BO), the MFA, and the GE are the Rumor Centrality (RC), the Degree Centrality (DC) and the Jordan Center (JC). 
We have also run the Dynamical Age (DA), but due to its overall poor performance and its computational complexity, we have omitted it from the plots.
Our selection of the methods loosely follows the methods surveyed in~\cite{Jiang2017}. Each curve shows the performance of one method for different values of the infection size, $2 \leq |O| \leq 300$.  Each point is an average over 500 infection paths rooted at random sources. To avoid an unreasonable computation time, we skip 
the optimal Bayes for the infected sets of size greater than 10. 
The BO curve serves as the benchmark for the best achievable performance. Note that even the optimal solution needs to output a large set to catch the source, signifying the inherent difficulty of the problem.

Rumor and Jordan centralities perform optimally on regular trees in Figure~\ref{fig:regular:tree},  as expected by the theory~\cite{Shah2011,Zhu2016}, although the network is not exactly an \emph{infinite} tree. Notice that RC, JC, and BO overlap for infection sizes not exceeding the depth of the tree. Degree centrality also appears as a close competitor in this figure. Moving to other networks, however, these popular methods do not perform better than random guessing. For all three, the expected relative rank is close to 0.5, even in a random tree. The plots in this section show that, despite their popularity, the RC and JC are quite unreliable for source recovery.

Among our proposed methods, MFA outperforms RC, JC, and DC in Figures~\ref{fig:ucsc68}, \ref{fig:uc64}, \ref{fig:wiki:vote}, \ref{fig:dc:sbm}. MFA finds the true source, on average, in its top 30\% guesses. 
 The networks with suprior MFA performance have highest transitivity (aka clustering coefficient) in Table~\ref{tab:stats}, that is, many triangles among triples of nodes. Transitivity has been studied extensively and it distinguishes human social networks from random trees and less cyclical networks, such as water distribution systems and traffic network. In this sense, MFA is suitable for rumor source detection in social networks.

GE is the global winner, except in regular trees (Figure~\ref{fig:regular:tree}). We were surprised that a greedy algorithm had such a widespread success. 
GE not only performs well in highly transitive networks, but it also outperforms RC, JC, and DC on random trees (Figure~\ref{fig:random:tree}) and less transitive networks (Figures~\ref{fig:internet}, \ref{fig:power}). 

\begin{wrapfigure}{r}{0.5\textwidth}
    \centering
    \includegraphics[trim={.9cm .2cm 1.1cm 1.2cm}, clip, width=.5\textwidth]{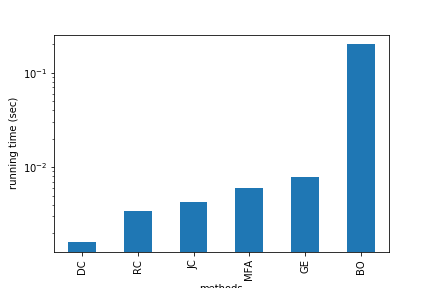}
    \caption{Runtime in seconds at infection size 10.} 
    \label{fig:uc64:lapse}
\end{wrapfigure}
Figure~\ref{fig:uc64:lapse} illustrates the runtimes (on the log scale) for a single run on the UC64 network, when the infection size is 10 (the maximum size for which the Bayesian results are available). 
Degree centrality is the fastest, followed by RC, JC, MFA and GE, all four having comparable speed, with RC and JC having a slight edge.
BO is about 10 times slower and its runtime grows exponentially with infection size.

Based on these results, we advocate for the use of GE as the main tool for identifying sources of epidemics, regardless of the network topology or the nature of  the epidemics (rumor propagation, disease contagion, etc.). MFA should be applied with caution. It is superior in social (transitive) networks, and attractive for its simplicity and the potential for parallelism. 



\bibliographystyle{unsrt}
\bibliography{refs3}

\newpage
\appendix
\section{Multi-source Extension}\label{sec:multi:source}
The inference problem discussed in Section~\ref{sec:bayes:infer} immediately extends to the multi-source situations. Consider the case were more than one independent source, denoted by $\bm I_*$, initiate the infection dynamics. Due to the Markovian nature of the dynamics, the infection path that leads to some set $I$ does not influence the value of $\rho_{I \to O}$. Hence, Proposition~\ref{prop:bayes} also describes the likelihood of the transition from the source set $\bm I_*$ to a snapshot $O$. 

If we know that there are $s$ original sources, e.g. $|\bm I_*|=s$, with a uniform prior on the patient zeros, the Bayesian solution would be characterized by the optimization 
\begin{align}\label{eq:map:multi}
I^*_\text{MAP} = \argmax_{I\subset O,\, |I|=s}\, \rho_{I \to O}
\end{align}
To compute this MAP estimate, we can still use the DP solution in Proposition~\ref{prop:bayes}, but we do not need to compute $\rho_{I \to O}$ for $|I| < s$.
Thus, the multi-source problem is in a sense ``easier'', especially when $s \approx |O|$, since one can terminate the recursion earlier (i.e., the case $s=1$ is the hardest).

\section{Proofs}
\subsection{Proof of Proposition~\ref{prop:bayes}}
	%
	Let us first recall a known fact about the exponential distribution:
	
	\begin{fact}\label{fact:indep:exponen}
		Let $T_i \sim \Expd(\beta_i)$ be a collection of independent exponential variables. Then,
		\begin{align*}
		\pr\Big(T_i < \min_{j \neq i} T_j\Big) = \frac{\beta_i}{ \sum_{j} \beta_j}.
		\end{align*}
	\end{fact}
	
	The forward programming~\eqref{eq:for:rec} is an application of the law of total probability in the following sense: The event that nodes in $O\setminus I$ are infected before any other node in $I^c$  splits into sub-events that each node in $O\setminus I$ is infected before those in $O^c$ and we have 
	\begin{align*}
	\rho_{I \to O} = \sum_{j \in O \setminus I} \rho_{I \to I \cup j} \cdot \rho_{I \cup j \to O} 
	\end{align*}
	where we have also used the Markov property of SI dynamics to split the probabilities on the RHS into the products.
	The ratio in~\eqref{eq:for:rec} corresponds to the transition probability from $I$ to $I\cup j$, that is $\rho_{I \to I \cup j}$. Indeed, given that $I$ is infected, we run exponential clocks $T_j \sim \Expd( \beta \vol(I,j))$ and the first to expire determines the next infected node. By Fact~\ref{fact:indep:exponen}, this happens for any node $j \in I^c$ with probability probability $\propto_j \beta \vol(I,j)$. Thus, 
	\begin{align*}
	\rho_{I \to I \cup j} = \frac{\beta \vol(I,j) }{\sum_{j'}\beta \vol(I,j')} = \frac{\vol(I, j)}{\vol(I, I^c)}.
	\end{align*}
	This proves the forward programming.
	%
	The backward programming, on the other hand, connects $\rho_{I \to O}$ to $\rho_{I \to O\setminus j}$ and is proved similarly. Basically, the event of visiting $O$ can be divided into sub-events based on the last node in $O$ that is infected. 
	%

\subsection{Proof of Proposition~\ref{prop:mean:field:lin:system}}\label{app:proof:mean:field}

We prove the following alternative expressions for  $S = (S_{jj'})^{|O| \times |O|}$ and $\zb = (z_j)^{|O|}$,
\begin{align*}
S_{jj'} &:= 
\begin{cases}
\deg_{O \setminus j'}(j) \cdot \deg_{O \setminus j}(j')   + \vol_{O \setminus \{j,j'\}}^{(2)}(j,j') & j \neq j' \\
2 \deg_O(j) \cdot [ \deg_O(j) + 1] & j = j'
\end{cases} \\[1ex]
z_j &:= \Big[\vol(O{\setminus j})  + 2 \vol\big((O{\setminus \,j})^c, O{ \setminus\, j}\big) \Big]\deg_O(j) \\ & \qquad 
+ 2 \vol\big(\adj_{O}(j), (O{\setminus \,j})^c\big).
\end{align*}
Here, $\deg_O(i) := \sum_{j \in O} A_{ij}$ is the degree of node $i$ in $O$,
and $\vol^{(2)}(i,j) := \sum_{r \in O} A_{ir} A_{rj}$ is the number of paths of length 2 between nodes $i$ and $j$ that pass through $O$.
It is not hard to verify that these expressions are equivalent to the matrix form presented in~\eqref{prop:mean:field:lin:system}.

Recall that $\vol(I, I^c) = \sum_{i,k} A_{ik} 1\{i \in I, k \notin I \}$ and similarity $\vol(I, j) = \sum_{r} A_{rj} 1\{r \in I \}$. Here, the indices, $i$, $k$ and $r$ run over all nodes in the network, i.e. $i,k,r \in [n]$. We have
\begin{align*}
(Q^T \rb)_j
&= \sum_{I\subset O} 1\{j\not\in I\}\, \vol(I, j)\cdot \vol(I, I^c) \\
&= \sum_{I \subset O \setminus \{j\}}\,  \vol(I, I^c) \cdot\vol(I, j) \\
&= \sum_{I \subset O \setminus \{j\}}\, \sum_{i,k,r} A_{ik} A_{rj} \, 1\{i \in I, \,k \notin I, \,r \in I\} \\
&= \sum_{i,k,r} A_{ik} A_{rj}  \gamma_{ikr}
\end{align*}
where the last equality follows by interchanging the order of summations and defining
\begin{align*}
\gamma_{ikr} := \sum_{I \subset O \setminus \{j\}}1\{i \in I, \,k \notin I, \,r \in I\} 
\end{align*}
If $ i$ or $r$ do not belong to $O\setminus\{j\}$, or $k  \in \{i,r\}$, then $\gamma_{ikr} = 0$. Thus, it what follows assume that $i,r \in O_{\setminus\, j} := O \setminus\{j\}$ and $k \notin\{i,r\}$. Then,
\begin{align*}
\gamma_{ikr} = 0
\begin{cases}
2^{|O|-4} & i \neq r,\; k \in   O_{\setminus\, j}\\
2^{|O|-3} & i=r, \;  k \in   O_{\setminus\, j}\\
2^{|O|-3} & i \neq r,\; k \notin   O_{\setminus\, j}\\
2^{|O|-2} &  i=r, \;  k \notin   O_{\setminus\, j}\\
\end{cases}
\end{align*}

To see the second equality, note that we are counting subsets of the set $O \setminus\{j\}$ (of cardinality $|O|-1$) that contain or exclude certain elements. For example, when $k,i,r$ are pairwise distinct, and $k \in  O \setminus\{j\}$, looking at the binary representation of $I$, we have two ones in the positions $i$ and $r$ and a zero in position $k$, and the rest of $|O|-1-3$ positions are free to be zero or one.

\smallskip
Let $\deg_{S}(i) = \sum_{j \in S} A_{ij}$ be the degree of node $i$ in $S$. We drop $S$ when $S = [n]$.
In what follows, $i$ and $r$ range over $O \setminus\{j\}$ (otherwise $\gamma_{ikr} = 0$). Also, condition $k \notin \{i,r\}$ can be replaced with $k \neq r$, since the $k\neq i$ is implicitly enforced by $A_{ik} = 0$ if $k = i$ (no self-loops). We have
\begin{align*}
(Q^T \rb)_j &= \sum_{i,r} \sum_{k \neq r} A_{ik} A_{rj} \big[ 2^{|O|-4}(1+1\{i=r\})1\{k \in   O_{\setminus\, j}\} \\ 
&+  2^{|O|-3}(1+1\{i=r\})1\{k \notin   O_{\setminus\, j}\}\big] \\
&= 2^{|O|-4} \sum_{i,r}  \deg_{O\setminus\{j,r\}}(i) A_{rj} (1+1\{i=r\}) \\
&+ 
2^{|O|-3}\sum_{i,r} \deg_{(O\setminus \,j)^c}(i) A_{rj} (1+1\{i=r\})    
\end{align*}
where in the second term, we used the fact that if $k \notin O_{\setminus\,j}$ then we automatically have $k \neq r$ since $r$ ranges over $O_{\setminus\,j}$. We have
\begin{align*}
\sum_{r}  \deg_{O\setminus\{j,r\}}(i) A_{rj} &= \sum_{r}  (\deg_{O_{\setminus j}}(i) - A_{ir}) A_{rj} \\
&= \deg_{O_{\setminus j}}(i) \deg_{O_{\setminus j}}(j) - \vol_{O_{\setminus j}}^{(2)}(i,j)
\end{align*}
where $ \vol_{O_{\setminus j}}^{(2)}(i,j) := \sum_{r \in O_{\setminus j}} A_{ir} A_{rj}$ is the number of paths of length two between $i$ and $j$ in $O_{\setminus j}$. Note that $ \vol_{O_{\setminus j}}^{(2)}(i,j) = \vol_{O}^{(2)}(i,j)$ and similarly $\deg_{O_{\setminus j}}(j) = \deg_{O}(j)$ since $A_{jj} = 0$. Thus,
\begin{align*}
\sum_{i,r} \deg_{O\setminus\{j,r\}}(i)\, A_{rj} \big( 1+1\{i=r\} \big) 
&= \sum_i \Big[ \deg_{O_{\setminus j}}(i) \deg_{O}(j) - \vol_{O}^{(2)}(i,j) +  \deg_{O_{\setminus j}}(i) A_{ij} \Big] \\
&= \sum_i \deg_{O_{\setminus j}}(i) \deg_{O}(j) \\
&= \vol(O_{\setminus j}) \deg_{O}(j) 
\end{align*}
where $\vol(O_{\setminus j}) = \vol(O_{\setminus j},O_{\setminus j})$ and the third equality follows since we have
\begin{align*}
\sum_{i\in A} \vol_{A}^{(2)}(i,j) = \sum_{i \in A} \sum_{r \in A}  A_{ir} A_{rj} = \sum_{r \in A} \deg_A(r) A_{rj}
\end{align*}
which was used with $A = O_{\setminus j}$. Similarly, we have
\begin{align*}
\sum_{i,r} \deg_{(O\setminus \,j)^c}(i) A_{rj} (1+1\{i=r\}) &= 
\sum_{i} \deg_{(O\setminus \,j)^c}(i)  \big (\deg_{{O \setminus j}}(j)+A_{ij} \big)  \\
&=\vol((O{\setminus \,j})^c, O{ \setminus\, j}) \,\deg_O(j) \\
&+ \vol(\adj_{O}(j), (O{\setminus \,j})^c)
\end{align*}
It follows that
\begin{align*}
(Q^T \rb)_j &=2^{|O|-4} \Big[ \vol(O{\setminus j}) \deg_{O}(j)  + 2 \vol\big((O{\setminus \,j})^c, O{ \setminus\, j}\big) \,\deg_O(j) + 2 \vol\big(\adj_{O}(j), (O{\setminus \,j})^c\big)\Big].
\end{align*}

\paragraph{Calculating $Q^TQ$} Let us first take $j\neq j'$. Then, similar to the previous argument,
\begin{align*}
(Q^T Q)_{jj'} &= \sum_{I\subset O\setminus \{j,j'\}}  \vol(I, j)\, \vol(I, j') \\
&=\sum_{I\subset O\setminus \{j,j'\}}  \sum_{i,r} A_{ij}\, A_{rj'}  1\{i \in I,\;r \in I\} \\
&= 
\sum_{i,r} A_{ij}\, A_{rj'} \beta_{ir}
\end{align*}
where we have defined
\begin{align*}
\beta_{ir} &:= \sum_{I\subset O\setminus \{j,j'\}} 1\{i \in I,\;r \in I\}  \\
&= 2^{|O|-4} 1\{i\neq r\} + 2^{|O|-3}   1\{i=r\}\\ 
&= 2^{|O|-4} \big(1+ 1\{i=r\}\big)
\end{align*}
assuming $i,r \in O \setminus \{j,j'\}$, otherwise $\beta_{ir} = 0$. Thus, restricting summations over indices $i,r \in O \setminus \{j,j'\}$
\begin{align*}
(Q^T Q)_{jj'} &= 2^{|O|-4}\Big[ \sum_{i,r} A_{ij}\, A_{rj'}  +  \sum_{i} A_{ij} A_{ij'}\Big] \\
&= 2^{|O|-4}\Big[ \deg_{O \setminus j'}(j) \deg_{O \setminus j}(j')   + \vol_{O \setminus \{j,j'\}}^{(2)}(j,j')\Big]. \\
\end{align*}

Now consider the case $j = j'$. Then,
\begin{align*}
(Q^T Q)_{jj} &= \sum_{I\subset O\setminus \{j\}}  \vol(I, j)^2 \\
&=\sum_{I\subset O\setminus \{j\}}  \sum_{i,r} A_{ij}\, A_{rj}  1\{i \in I,\;r \in I\} \\
&= 
\sum_{i,r} A_{ij}\, A_{rj} \,2^{|O|-3} \big( 1 + 1\{i=r\} \big),
\end{align*}
assuming $i,r \in O\setminus j$. It follows that
\begin{align*}
(Q^T Q)_{jj} &= 2^{|O|-3} \Big[ \sum_{i,r} A_{ij}\, A_{rj} + \sum_{i} A_{ij} \Big] \\
&=  2^{|O|-3} \big[ \deg_{O}(j)^2 + \deg_{O}(j)].
\end{align*}

\end{document}

%% file: defs.tex
\newcommand\ub{\bm u}
\newcommand{\ih}{\bm {\hat  i}}
\newcommand{\sigh}{\bm {\hat  \sigma}}
\newcommand{\is}{\bm {i}_*}

\DeclareMathOperator{\vol}{vol}
\DeclareMathOperator{\adj}{adj}
\renewcommand{\deg}{d}
\DeclareMathOperator{\Expd}{Exp}

\newcommand{\rb}{\bm r}
\newcommand\xb{\bm x}
\newcommand\bb{\bm b}
\newcommand\bbh{\widehat \bb}
\newcommand{\one}{\mathbf{1}}

\newcommand{\zb}{\bm z}
\newcommand{\vb}{\bm v}